# Hybrid Automatic Neighbor Relations for 5G Wireless Networks

Ali Görçin*[†], Nicolae Cotanis[‡]
*Department of Electronics and Communications Engineering, Yıldız Technical University, İstanbul, Turkey
[†]Informatics and Information Security Research Center (BİLGEM), TÜBİTAK, Kocaeli, Turkey
[‡]George Mason University, Electrical & Computer Engineering Department, 4400 University Drive, Fairfax, VA, US
Emails: agorcin@yildiz.edu.tr, ncotanis@gmu.edu

*Abstract*—5G wireless networks aim to achieve seamless mobile broadband services, machine to machine and ultra-reliable low latency communications. These goals require massive improvements in terms of capacity, reliability, latency reduction, and scalability. Besides software defined networks and network function virtualization, self organizing networks (SON) is another enabling technology which makes 5G feasible. Automatic neighbor relations (ANR) is a prominent component of SON which has direct impact on reliability, scalability, and capacity of the wireless networks. When the current state of standardization and implementation of ANR functionality is considered, it is seen that a new approach is needed for ANR to comply with 5G requirements. Thus, in this paper a comprehensive hybrid ANR (H-ANR) architecture which leads to the management of relations utilizing an extensive set of universal performance criteria, configuration metrics, and measurements is proposed. The H-ANR functionality ranks the neighbors not based on instantaneous information but on historic values of the performance criteria and other metrics. The attributes of distributed ANR (D-ANR) module is also managed by H-ANR to prevent the repetitive addition of unwanted neighbors via D-ANR or X2 interface. Performance of the proposed H-ANR functionality is also compared with the case when only D-ANR is available on a Long Term Evolution Advanced network.

## I. Introduction

When compared to the currently available wireless communications technologies, 5G systems imply more sophisticated networks that can provide seamless connectivity to users while supporting high throughput; low latency in term of data delivery and flexibility in terms of scalability are needed to accommodate a large number of users and devices which should be ubiquitously connected. In order to satisfy such requirements, vendors and mobile radio network operators face challenging and complex operational tasks. To this end, self organizing networks (SON) are the technical solution to achieve management of complex optimization tasks and thus to provide operating expense (OPEX) reduction and revenue protection while maintaining the preset quality of service (QoS) benchmarks for wireless networks.

Automatic Neighbor Relations (ANR) is a SON function which is introduced to eliminate the need to manually manage neighbor relations (NRs) in cellular wireless networks. ANR can be activated at self-configuration stage to populate initial neighbor relation table (NRT), at self-optimization stage to optimize the NRT, and at self-healing stage to manage inactive NRs. Total network performance improvements are achieved via ANR operations because the cells under a given base station (BS) has optimized and up-to-date tables at every step of operations. It is shown that optimized NRTs means improvements in terms of handover (HO) timings, number of successful HOs and reduction in drop call rates which occur due to missing NRs [1]. Therefore ANR functionality has significant impact on the network performance in terms of capacity, latency and scalability which are the key optimization criteria for 5G networks [2].

Results on ANR research and current status are presented in [3]-[15]. When 5G related ANR works are considered, it is seen that [16] introduced automatic BS relations approach to detect neighbor cells in 5G systems. Downlink, uplink, and link failure procedures are described along with virtual beam relation establishment. [17] proposed utilization of network function virtualization for operating ANR to make it as flexible as possible to adapt the dynamic environment of 5G networks. On the other hand, ANR concepts are included in long term evolution (LTE) standards starting from the first release of the technology (Release 8), and its scope is expanded with subsequent 3GPP documentation releases, *i.e.*, TS 32.511, TS 36.300, TS 32.500. These documents regulate ANR operations in a broad sense indicating ANR feature should be able to add relations to NRT or remove them from NRT, it should have relation and X2 connection whitelisting and blacklisting, HO allowance and prohibition options. More importantly, the only detailed ANR operations is the discovery of new neighbors via user equipment (UE) measurements for intra-frequency and inter-frequency cases which is a part of distributed ANR (D-ANR) architecture. Thus the scope of ANR definitions falls short when the requirements of current and future wireless systems are taken into account.

In this paper a comprehensive hybrid ANR (H-ANR) architecture is proposed considering the requirements of 5G systems which will keep LTE-A at the core of wireless cellular network. The introduced architecture is also applicable to current networks and includes NR ranking, blacklisting and whitelisting functionalities which utilize an extensive set of universal performance criteria, configuration metrics, and measurements. These functionalities are applicable to inter-vendor/heterogeneous network (HetNet) scenarios. Furthermore, D-ANR attributes of BSs are managed to optimize automatic intra-frequency neighbor detection process. eX2 blacklisting and whitelisting policy based on cell to cell or transmission/reception points (TRP) to TRP relations [16] between two BSs are also envisioned and introduced for 5G systems. Optimization of NRT which includes removal of poorly performing neighbors is conducted by ranking NRs with a universal ranking equation. New NRs are introduced utilizing historical performance of NRs and number of available vacancies.

## II. COMPARISON OF ANR ARCHITECTURES

ANR can operate in centralized mode (C-ANR), in distributed mode, or in hybrid mode. These three architectures are illustrated in Fig. 1 for a combination of 5G BSs with TRPs, 4G eNodeBs, and 3G UTRAN NodeBs.

### A. Distributed ANR

In case of D-ANR, BSs add their intra-frequency LTE neighbors automatically monitoring signaling messages over Uu and X2. Thus in D-ANR neighbor detection process is typically fast, as it is deployed close to the radio network. However the process influences fewer cells per-algorithm because D-ANR module works per BS basis as shown in Fig. 1a. Moreover, If the UE which detects physical cell identifier does not have the EUTRAN cell global identifier (ECGI) reading capability, distributed ANR may not be able to update the NRT with detected cell unless ECGI information is provided by the central mobility management entity (MME).

When the current implementation of D-ANR for Long Term Evolution Advanced (LTE-A) is considered, even though there are some performance management (PM) counters which mark additions or removals of NRs, there is no information stored about *which* neighbors are added or removed. Therefore, same neighbor can be added and removed repeatedly without any measure taken against such behavior. Furthermore, there is no comprehensive removal policy in D-ANR. Neighbors are removed from NRT when there is certainly no HO to that cell for a predefined period of time. In such a setting, neighbors which are far away with very low HO share overpopulate NRTs. Even though some measures such as temporary neighbor monitoring modes are taken solely based on HO share, timer based removal process is still very slow and cumbersome. Moreover, even though D-ANR can detect neighbors and add them automatically to NRTs of cells, it still requires an extensive manual configuration and user supervision process. Parameters such as:

- reference signal received power (RSRP) and quality (RSRQ) thresholds along with D-ANR A3 and A5 threshold offsets for adding a neighbor,
- maximum and minimum numbers of UEs, increasing and decreasing numbers of UEs to detect a neighbor,
- removal periods for relations with no handover activity for each cell under each BS

should be configured beforehand, manually. If this process is not conducted carefully, D-ANR operation can significantly effect network and UE resource consumption. Besides, each UE can scan only one frequency for a predefined period of time for ANR purposes. This is a limiting factor for detecting new neighbors through D-ANR. Current implementation of D-ANR does not have any blacklisting and whitelisting policy. Such policies for D-ANR are not proposed in the literature either. Therefore D-ANR does not conduct blacklisting and whitelisting of cells in the NRT. On the other hand, X2 is a logical interface which leads to direct connection and communication of eNodeBs in LTE systems. Currently, mobility and load management or general error reporting are primitively handled through X2 connections. If properly configured, reciprocal neighbor additions are also conducted through X2 connections between the cells of different eNodeBs. In the 5G wireless networks, the functionality of interfaces such as eX2 and eS1 are expected to increase not only the scalability of wireless networks but also robustness of M2M communications. D-ANR currently does not have any (e)X2 blacklisting or (e)X2 whitelisting policy, therefore cases such as

- removal of unnecessary (e)X2 connections with BSs with no cells or TRPs as an NR under managed BS and (e)X2 blacklisting in case of repetitive attempts of unnecessary connections,
- (e)X2 blacklisting of a BS that belongs to another unwanted PLMN,
- (e)X2 whitelisting of a BS with relations that have high key performance indicators (KPIs),

are not handled by D-ANR. Finally, for inter-frequency and inter-RAT NRs, D-ANR requires significant assistance to start automatic neighbor detections including creation of multiple managed objects.

### B. Centralized ANR

In contrast with D-ANR, C-ANR can oversee multiple cells under different BSs all together as shown in Fig. 1b. Thus, C-ANR can jointly optimize NRTs of multiple BSs. C-ANR adds NRs to cells based on centralized configuration data and KPIs. ECGI is acquired from centralized database, instead of UE measurements. However, C-ANR may detect potential neighbors slower when compared to D-ANR since it does not have real-time intra/inter RAT and intra/inter frequency NRT update functionalities that D-ANR possesses.

When only traditional C-ANR is available, all ANR operations depend on centralized information and in case of a missing neighbor, analysis of centralized data can take some time and additions of new neighbors can take some more time. Moreover, traditional C-ANR operates with instantaneously available data. There is no performance monitoring of the neighbors or decision making based on the performance history. Such decisions can lead to deletion of good NRs which can perform poor in shorter time intervals and addition of short-term good performing neighbors which performs poorly in the long run. On the other hand, when traditional C-ANR operates along with D-ANR, C-ANR does not consider operations of D-ANR and focuses on the optimization of NRT as an independent entity. C-ANR can remove some poorly performing neighbors depending on some predefined criteria, however these neighbors can be brought back by D-ANR. Therefore some ping-pong effect can occur between C-ANR and D-ANR. This is also a side effect of the fact that traditional C-ANR does not manage or configure D-ANR attributes for optimized automatic neighbor detection.

In general traditional C-ANR has only NR blacklisting and whitelisting operations based on a small set of KPIs excluding RSRP and RSRQ indicators that can be extracted from cell traces recorded via UEs. As in the case of D-ANR, C-ANR does not also include policies for (e)X2 blacklisting, (e)X2 whitelisting or detection of unwanted NRs or BSs.

### C. Hybrid ANR

In the general H-ANR architecture, algorithms are deployed at the BSs and a set of open interfaces are provided to implement a control algorithm that operates on a wider geographic area, on a longer time scale. H-ANR can jointly

- oversee network elements from different vendors, managing the problems that can occur due to the incompatibility of different technologies in a network,

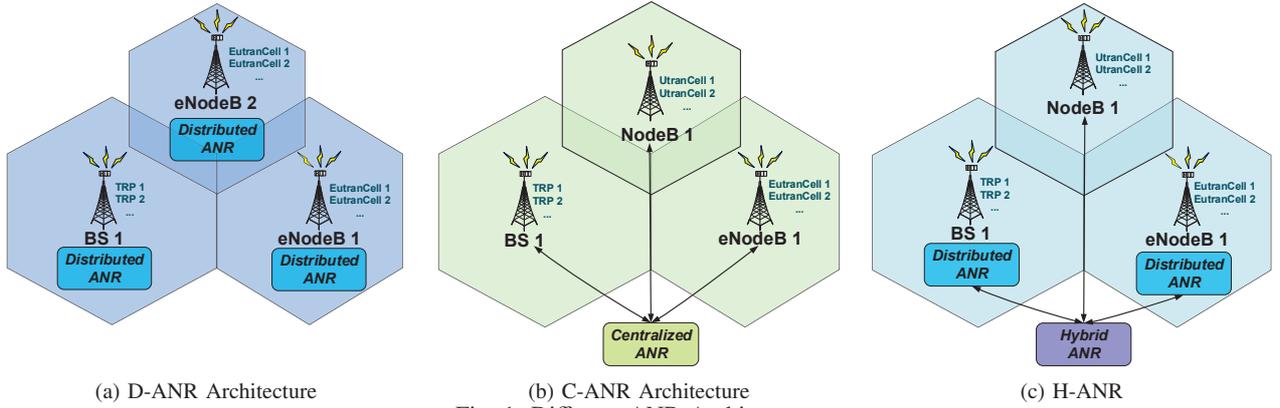

Fig. 1: Different ANR Architectures

(a) D-ANR Architecture  (b) C-ANR Architecture  (c) H-ANR

- provide an initial configuration for the D-ANR, since D-ANR may not perform well until there is sufficient traffic to make instantaneous NR additions,
- optimize BSs when D-ANR feature is deactivated, not licensed or available as shown in Fig. 1b,
- track and monitor live performance data from the D-ANR modules and use this data to feed into central evaluation and optimization process of ANR, as well as other SON modules,
- conduct traditional C-ANR functionalities.

Even though H-ANR is an optimum comprise between D-ANR and C-ANR, H-ANR operations have higher complexity which cause increased control plane signaling. On the other hand, its current scope is not sufficient to comply with 5G requirements described in Section I, especially with dynamic HetNet optimization which strives for very low latency under high scalability.

## III. H-ANR Architecture for 5G Wireless Networks

Table I summarizes the differences of C-ANR, D-ANR and proposed H-ANR architectures. H-ANR introduces a comprehensive ANR management policy which is comprised of:

- building the NRT from scratch if it is empty and tracking additions and removals of NRs one by one to prevent repetitive additions and removals of same neighbors,
- management of D-ANR attributes to optimize automatic intra-frequency neighbor detection, and to prevent repetitive additions of poorly performing NRs by D-ANR,
- NR blacklisting, whitelisting, and PLMN blacklisting depending on an extensive set of performance and configuration metrics and cell traces,
- BS level (e)X2 blacklisting and whitelisting depending on cell to cell relations between two BSs,
- automatized intra-frequency, inter-frequency and inter-RAT relation additions compatible to D-ANR operations using centralized information collection and evaluation capability,
- optimization of NRT by ranking the NRs, removal of poorly performing neighbors and introducing new relations based performance, configuration management and cell trace data. Removals and additions are conducted in a compatible way to avoid conflict with the D-ANR operations.

Therefore proposed H-ANR architecture provides fundamental functions which utilize an extensive set of universal performance criteria, configuration metrics, and measurements such that these functions are applicable to high demanding inter-vendor/HetNet 5G networks. Due to the space limitations paper will detail D-ANR attribute management and NRT optimization features while providing general information about H-ANR functionality.

### A. Stages of H-ANR Operations

H-ANR has four stages of operations which are conducted periodically in the sequence as illustrated in Fig. 2 assuming that in 5G networks interface such as *Itf-N* will provide data stream from network, reliably. Please note that H-ANR runs with the period of *anr_run_time* in a previously set *time_window*. *run_counter* is incremented every time H-ANR starts a new cycle and H-ANR operations cease as the *time_window* is completed. Functionalities of the blocks in Fig. 2 are presented subsequently.

*1) Parameter Management:* As the H-ANR starts operating, at the beginning of each cycle, additions to and removals from the NRT are detected and internal tables are updated. NR additions and removals can be conducted by H-ANR, D-ANR, or the operator. New NRs can also be added through X2. Thus H-ANR tracks all these changes by keeping track of additions and removals of relations in the NRT which leads to the monitoring of multiple additions and removals of same NR in different period of times by different controllers. Therefore it becomes possible to track repetitive actions in NRT and to take preventive measures against such behavior during the maintenance of it. Information tracking is also utilized while making whitelisting and blacklisting decisions for NRs. Example table for NR additions can be seen in Fig. 3. Parameter management also checks if NRT is empty or not at each run. NRT could be emptied for different reasons *e.g.*, if the cell or BRS had a reset, due to some software or hardware glitches. In such a case, NRT is filled up with the geographically close neighbors based on centralized configuration management data.

*2) D-ANR Attribute Management:* Since operators can opt in or out running D-ANR functionality taking their OPEX

TABLE I: Comparison of ANR Architectures for Next Generation Wireless Networks

| Feature | H-ANR | D-ANR | C-ANR |
|---|---|---|---|
| Automated intra-frequency NR Management | Yes | Yes | Yes |
| Automated inter-RAT NR Management | Yes | No | No |
| Automated D-ANR Attribute Management | Yes | No | No |
| Automated Relation Blacklisting | Yes | No | Yes |
| Automated Relation Whitelisting | Yes | No | No |
| Automated X2 Blacklisting | Yes | No | No |
| Automated X2 Whitelisting | Yes | No | No |
| X2 Connection Management | Yes | No | No |
| Relation Tracking | Yes | No | No |
| Relation Deletion Decision Resolution | Seconds | Days | Seconds |
| Relation Deletion Policy | Mobility KPIs, cell traces, configuration data | Number of HOs | Distance and HO success rate |
| Relation Deletion Decision Policy | Based on historical data, memory | Memoryless | Memoryless |
| Operation Type | Multiple base stations | Per cell under each base station | Multiple base stations |
| Cooperation with available D-ANR | Yes | N/A | No |

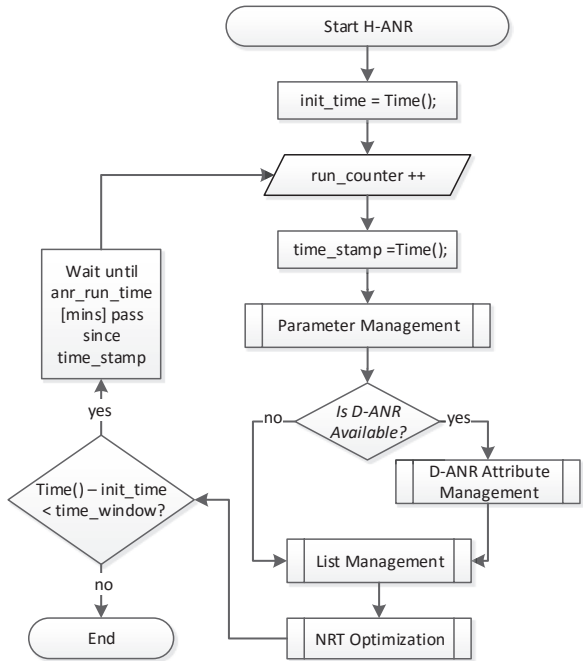

Fig. 2: Flowchart of H-ANR functions

Fig. 3: NR additions tracking table. Database IDs of target cells aids to the detection of repetitive behavior since NR IDs can be different at every addition event, even though they represent the same target cell.

requirements into consideration, D-ANR does not necessarily have to be available at each BS in a 5G network as indicated in Fig. 1c. Therefore, after the checksum of NRT, H-ANR checks if the D-ANR functionality is available and active or not over the BS under optimization. If D-ANR is available, H-ANR controls D-ANRs behavior by automatically modifying D-ANR control parameters to prevent addition of unwanted neighbors (*e.g.*, far away over-shooters) or repetitive neighbor additions and removals. For instance, H-ANR monitors and modifies attributes such as RSRP and RSRQ thresholds to prevent addition of poorly performing neighbors by D-ANR. In this context, assuming that the number of successful HOs to NR $r$ from the source cell (SC) in *anr_run_time*($i$) (which is between $(i-1)^{th}$ and $i^{th}$ runs of H-ANR) is $S_{HO}(r,i)$, total number of successful HOs from the SC is given by

$$TS_{HO} = \sum_{r=1}^{nr} S_{HO}(r,i) \qquad (1)$$

where $nr$ is the total number of NRs in the NRT of SC. Therefore *normalized* successful HOs from SC to NR $r$ can be calculated by

$$NS_{HO}(r,i) = \frac{S_{HO}(r,i)}{TS_{HO}}. \qquad (2)$$

and since H-ANR depends on historic data over an *observation_window*, the average normalized successful HOs from the SC to NR $r$ becomes

$$A\_NS_{HO}(r) = \mathbb{E}\{NS_{HO}(r)\} = \frac{1}{W}\sum_{i=1}^{W} NS_{HO}(r,i)$$

where $W$ is the *observation_window* size. During *anr_run_time*($i$) each UE connected to SC can conduct multiple RSRP measurements on relations, thus the averaged RSRP measurements for NR $r$ can be written as

$$RSRP(r,i) = \sum_{u=1}^{U} RSRP(r,i,u) \qquad (3)$$

where $U$ denotes the total number of RSRP measurements taken from the NR $r$ by all UEs connected to SC. On the other hand, assuming that the maximum averaged RSRP measurement is $max\_RSRP = \max\{RSRP(r,i)\}$ during *anr_run_time*($i$), *normalized* RSRP for NR $r$ becomes

$$NRSRP(r,i) = \frac{RSRP(r,i)}{max\_RSRP}, \qquad (4)$$

**Algorithm 1** D-ANR Attribute Management

1: **variables**
2:    $HO\_MIN\_THR$: Minimum HO share NR should satisfy
3:    $RP\_THR$: Minimum RSRP level NR should satisfy
4:    $CELL\_RSRP\_THR$: Internal D-ANR RSRP threshold for NR additions
5:    $bad\_additions$: Bad NR additions by D-ANR
6: **end variables**
7: **procedure** RSRP THRESHOLD MANAGEMENT
8:    **for** all cells and TRPs under BS **do**
9:      **for** all new NRs under cell or TRP **do**
10:        **if** (Created by = D-ANR) **then**
11:          **if** ($A\_NS_{HO}(r) < HO\_MIN\_THR$) **then**
12:            **if** ($A\_NRSRP(r) < RP\_THR$) **then**
13:               $bad\_additions(run\_counter)$
14:               $\leftarrow bad\_additions(run\_counter) + 1$
15:            **end if**
16:          **end if**
17:        **end if**
18:      **end for**
19:    **end for**
20:    **if** ($run\_counter = 1$ & $bad\_additions > 0$) **then**
21:       $CELL\_RSRP\_THR \leftarrow CELL\_RSRP\_THR + 1$
22:    **else if** ($bad\_additions(run\_counter) >$
23:    $bad\_additions(run\_counter - 1)$) **then**
24:       $CELL\_RSRP\_THR \leftarrow CELL\_RSRP\_THR + 1$
25:    **end if**
26: **end procedure**

while the average normalized RSRP measurements for NR $r$ is written as

$$A\_NRSRP(r) = \mathbb{E}\{NRSRP(r)\} = \frac{1}{W}\sum_{i=1}^{W} NRSRP(r,i).$$

and the algorithm to control the RSRP threshold of D-ANR is given in Algorithm 1. RSRQ threshold is also adjusted with the same approach.

*3) List Management:* If unwanted NRs continue to come back either through X2 or D-ANR after D-ANR attributes are optimized, blacklisting is activated. H-ANRs blacklisting policies isolate such neighbors from further relation establishment. H-ANR also has a whitelisting policy which prevents the removal of consistently well-performing neighbors from NRT. H-ANR provides an extensive (e)X2 blacklisting policy to ban poor performing BSs or to prevent unwanted BSs to establish (e)X2 connections. (e)X2 whitelisting is applied when all cells between two BSs are very highly ranked at all NRTs of all cells, mutually.

*4) NRT Optimization:* Decisions on which NRs to delete, keep and add are made in this stage. H-ANR ranks the NRs based on the ranking equation which comprised from an extensive set of metrics. Again, assuming the number of HO execution attempts to $r$ from the SC in $anr\_run\_time(i)$ is $A_{HO}(r,i)$, *relatively* successful HOs can be calculated by

$$RS_{HO}(r,i) = \frac{S_{HO}(r,i)}{A_{HO}(r,i)}, \quad (5)$$

and average relative successful HOs is then

$$A\_RS_{HO}(r) = \mathbb{E}\{RS_{HO}(r)\} = \frac{1}{W}\sum_{i=1}^{W} RS_{HO}(r,i).$$

On the other hand, if total number of attempted HO preparations to $r$ is given by $AP_{HO}(r,i)$, and successful HO preparations to $r$ is given by $SP_{HO}(r,i)$, *relative* successful HOs preparations can be written as

$$PS_{HO}(r,i) = \frac{SP_{HO}(r,i)}{AP_{HO}(r,i)}, \quad (6)$$

and average relative successful HO preparations becomes

$$A\_PS_{HO}(r) = \mathbb{E}\{PS_{HO}(r)\} = \frac{1}{W}\sum_{i=1}^{W} PS_{HO}(r,i).$$

Assuming that the distance of farthest away NR from the SC is

$$max\_dist = \max_{i}\{dist(i)\},$$

the *normalized* distance of $r$ from the SC is then

$$N\_DIST(r) = \frac{dist(r)}{max\_dist}, \quad (7)$$

and averaged normalized RSRQ value is $A\_NRSRQ(r)$ and can be calculated in a similar manner with that of $A\_NRSRP(r)$. The ranking of $r$ is given by

$$Rank(r) = A\_NS_{HO}(r) + (A\_RS_{HO}(r) \times A\_PS_{HO}(r)) + \quad (8)$$
$$A\_NRSRP(r) + A\_NRSRQ(r) - N\_DIST(r).$$

Note that, instead of RSRP and RSRQ, received signal chip power and $E_c/N_0$ are employed for UTRAN, and received signal strength indicator is utilized for GERAN neighbors. The NRs to delete are detected based on cumulative sum detection method which discovers significant changes through the bottom of NRT utilizing the soft ranking values of NRs calculated with eqn. 8.

IV. MEASUREMENT RESULTS

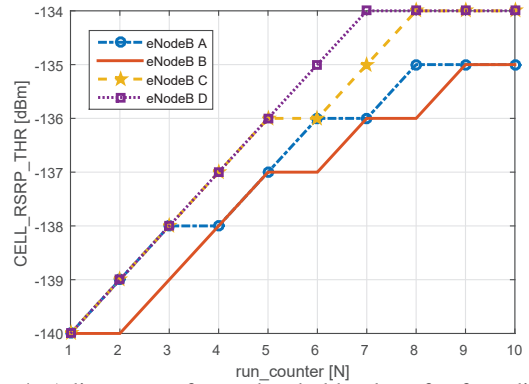

Fig. 4: Adjustment of rsrp threshold values for four different eNodeBs which are mutually exclusive in terms of coverage area.

The proposed H-ANR architecture is implemented over the LTE-A subnetwork of a first tier service provider in a very large urban city. The cluster that H-ANR functions run was total of 667 eNodeBs with 3542 eUtran cells. Fig. 4 shows how the $CELL\_RSRP\_THR$s for each eNodeB is adjusted by H-ANR according to algorithm 1 so that D-ANR does not add bad neighbors to NRT anymore. Due to the differences in the operational environment, each eNodeB has different threshold settling time. Fig. 5 also shows how $CELL\_RSRQ\_THR$s are adjusted with a similar approach to RSRP thresholds.

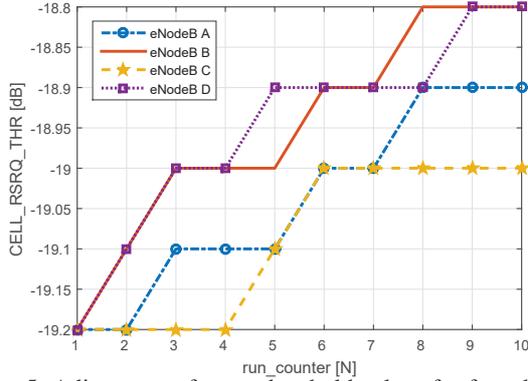

Fig. 5: Adjustment of rsrpq threshold values for four different eNodeBs which are mutually exclusive in terms of coverage area.

Table II lists relations which are added either by D-ANR or X2, and kept in the NRTs but removed by H-ANR after it is activated. The removal decision is made based on the ranking equation (eqn. 8). Many of the NRs removed depending on relations' consistent poor performances over the *observation_window*. Note that statistically the majority of relations imply $A\_NS_{HO}(r) < 0.01$ and very rarely $0.01 < A\_NS_{HO}(r) \leq 0.02$. The average distance of removed NRs are also given in the last column for each cell. For consistency results are provided for the same eNodeBs which were utilized for D-ANR attribute management.

TABLE II: NRs removed from the NRTs of cells of eNodeBs after H-ANR is activated and operated for $run\_counter = 10$.

| eNodeB # | Cell # | X2 | D-ANR | Average Distance |
|---|---|---|---|---|
| eNodeB A | Cell 1 | - | 5 | 15, 86 km |
| | Cell 2 | - | 9 | 9, 66 km |
| | Cell 3 | 1 | 6 | 9, 71 km |
| | Cell 4 | 2 | 19 | 10, 47 km |
| | Cell 5 | - | 6 | 13, 33 km |
| | Cell 6 | - | 4 | 16, 75 km |
| eNodeB B | Cell 1 | 2 | 4 | 11, 66 km |
| | Cell 2 | 2 | 3 | 10, 80 km |
| | Cell 3 | 1 | - | 7, 63 km |
| | Cell 4 | 1 | - | 8, 27 km |
| | Cell 5 | - | 1 | 7, 16 km |
| | Cell 6 | - | - | - |
| eNodeB C | Cell 1 | 4 | - | 11, 25 km |
| | Cell 2 | 10 | - | 12, 30 km |
| | Cell 3 | 6 | 6 | 10, 41 km |
| | Cell 4 | 3 | 13 | 8, 97 km |
| | Cell 5 | 2 | 2 | 8, 50 km |
| | Cell 6 | - | 11 | 8, 72 km |
| eNodeB D | Cell 1 | 16 | - | 10, 69 km |
| | Cell 2 | 7 | - | 7, 32 km |
| | Cell 3 | 6 | 2 | 8, 37 km |
| | Cell 4 | 1 | - | 9, 84 km |
| | Cell 5 | 14 | 4 | 8, 55 km |
| | Cell 6 | - | - | - |

V. CONCLUSION

In this study an H-ANR architecture is proposed to satisfy the dynamic conditions of next generation 5G wireless networks. In contrary to the available C-ANR and D-ANR implementations, with its four stage operations H-ANR tracks the NR additions and removals one by one to prevent repetitive behavior, NRs are ranked, blacklisted and whitelisted considering an extensive set of metrics, D-ANR attributes are managed, and eX2 blacklisting and whitelisting is conducted as a part of the ANR optimization process. The measurements showed that D-ANR attributes should be optimized per eNodeB to prevent unnecessary additions by D-ANR. Poor performing NRs can also populate NRTs via X2. H-ANR provides solutions to these problems by adjusting D-ANR thresholds and utilizing a universal ranking equation.

REFERENCES

[1] Harri Holma, and Antti Toskala, "LTE for UMTS: Evolution to LTE-Advanced", John Wiley & Sons, 2011.
[2] J. Pérez Romero, O. Sallent, C. Ruiz, A. Betzler, P.S. Khodashenas, S. Vahid, K.M. Nasr, B. Abubakar, A. Whitehead, L. Goratti, "Self-X in SESAME", In *Proc. European Conference on Networks and Communications, EUCNC 2016*, Athens, Greece, 27-30 June 2016.
[3] H. Olofsson, S. Magnusson, and M. Almgren, "A Concept For Dynamic Neighbor Cell List Planning In A Cellular System", In *7th IEEE International Symposium on Personal, Indoor and Mobile Radio Communications, PIMRC'96.*, Taipei, Taiwan, 18-18 Oct. 1996.
[4] D. Soldani, and I. Ore, "Self-optimizing Neighbor Cell List For UTRA FDD Networks Using Detected Set Reporting," In *Proc. IEEE Vehicular Technology Conference, VTC2007-Spring*, Dublin, Ireland, 22-25 April 2007.
[5] J. Baliosian, and R. Stadler, "Distributed Auto-configuration Of Neighboring Cell Graphs In Radio Access Networks," *IEEE Transactions on Network and Service Management*, vol. 7, no. 3 pp. 145-157, 2010.
[6] F. Parodi, M. Kylvaja, G. Alford, J. Li, and J. Pradas, "An Automatic Procedure For Neighbor Cell List Definition In Cellular Networks", In *Proc. IEEE International Symposium on World of Wireless, Mobile and Multimedia Networks*, Helsinki, Finland, 18-21 June 2007.
[7] M. Amirijoo, P. Frenger, F. Gunnarsson, H. Kallin, J. Moe, and K. Zetterberg, "Neighbor Cell Relation List And Measured Cell Identity Management In LTE", In *Proc. IEEE Network Operations and Management Symposium, NOMS 2008*, Salvador, Bahia, Brazil, 7-11 April 2008.
[8] D. Huang, X. Wen, Bo Wang, W. Zheng, and Y. Sun, "A Self-optimising Neighbor List With Priority Mechanism Based On User Behavior", In *Proc. IEEE International Colloquium on Computing, Communication, Control, and Management, CCCM ISECS*, Sanya, China, 8-9 Aug. 2009.
[9] C. M. Mueller, H. Bakker, and L. Ewe, "Evaluation Of The Automatic Neighbor Relation Function In A Dense Urban Scenario," In *Proc. IEEE Vehicular Technology Conference, VTC2011-Spring*, Budapest, Hungary, 15-18 May 2011.
[10] Y. Watanabe, Y. Matsunaga, K. Kobayashi, H. Sugahara, and K. Hamabe, "Dynamic Neighbor Cell List Management For Handover Optimization in LTE," In *Proc. IEEE Vehicular Technology Conference, VTC2011-Spring*, Budapest, Hungary, 15-18 May 2011.
[11] A. Dahlén, A. Johansson, F. Gunnarsson, J. Moe, T. Rimhagen, and H. Kallin, "Evaluations of LTE Automatic Neighbor Relations," In *Proc. IEEE Vehicular Technology Conference, VTC2011-Spring*, Budapest, Hungary, 15-18 May 2011.
[12] Y.Li, Li Ji, and Li Yang, "Automatic Neighbor Relation Penetration Probability Prediction," In *Proc. IEEE Vehicular Technology Conference, VTC2012-Fall)*, Québec City, Canada, 3-6 September 2012.
[13] M. Taneja, V. Bangalore, G. Garuda, M. Nallathambi, and S. Gupta, "Policy Based Automatic Neighbor Relation management For Small Cell Networks," In *Proc. IEEE International Conference on ICT Convergence, ICTC*, Jeju Island, Korea (South), 14-16 Oct. 2013
[14] D. Duarte, P. Vieira, A. Rodrigues, N. Oliveira, and L. Varela, "Neighbour List Optimization For Real LTE Radio Networks," In *Proc. IEEE Asia Pacific Conference on Wireless and Mobile*, Bali, Indonesia, 28-30 Aug. 2014.
[15] M. Vondra, and Z.Becvar, "Distance-based Neighborhood Scanning For Handover Purposes In Network With Small Cells," In *IEEE Transactions on Vehicular Technology*, vol. 65, no. 2 pp. 883-895, 2016.
[16] P. Ramachandra, et al., "On Automatic Establishment Of Relations In 5G Radio Networks", In *27th IEEE International Symposium on Personal, Indoor and Mobile Radio Communications, PIMRC'16.*, Valencia, Spain, 4-8 Sept. 2016.
[17] Y. Shin and S. Kim, "Virtualized ANR to Manage Resources for Optimization of Neighbour Cell Lists in 5G Mobile Wireless Networks," In *Mobile Information Systems*, vol. 2017, Article ID 9643401, 2017. doi:10.1155/2017/9643401